# Fast spin information transfer between distant quantum dots using individual electrons

B. Bertrand[1,2], S. Hermelin[1,2], S. Takada[1,2,3], M. Yamamoto[3,4], S. Tarucha[3,5], A. Ludwig[6], A. D. Wieck[6], C. Bäuerle[1,2], T. Meunier[1,2]

[1] *Université Grenoble Alpes, Institut NEEL, F-38042 Grenoble, France.*

[2] *CNRS, Institut NEEL, F-38042 Grenoble, France.*

[3] *Department of Applied Physics, University of Tokyo, Tokyo, 113-8656, Japan.*

[4] *PRESTO-JST, Kawaguchi-shi, Saitama 331-0012, Japan.*

[5] *RIKEN Center for Emergent Matter Science (CEMS), 2-1 Hirosawa, Wako-Shi, Saitama 31-0198, Japan.*

[6] *Lehrstuhl für Angewandte Festkörperphysik, Ruhr-Universität Bochum, Universitätsstrasse 150, 44780 Bochum, Germany.*

**Transporting ensembles of electrons over long distances without losing their spin polarization is an important benchmark for spintronic devices. It requires usually to inject and to probe spin polarized electrons in conduction channels using ferromagnetic contacts[1,2] or optical excitation[3-5]. Parallel to this development, an important effort has been dedicated to the control of nanocircuits at the single electron level. The detection and the coherent manipulation of the spin of a single electron trapped in a quantum dot are now well established[6-8]. Combined with the recent control of the displacement of individual electrons between two distant quantum dots[9,10], these achievements permit to envision the**

**realization of spintronic protocols at the single electron level. Here, we demonstrate that spin information carried by one or two electrons can be transferred between two quantum dots separated by a distance of 4 µm with a classical fidelity of 65 %. We show that it is presently limited by spin flips occurring during the transfer procedure prior to and after the electron displacement. Being able to encode and control information in the spin degree of freedom of a single electron while being transferred over distances of a few microns on nanosecond timescales paves the way towards "quantum spintronics" devices where large scale spin-based quantum information processing could be implemented.**

To protect the information stored in a single electron spin, the electron has to be isolated from the other electrons of the nanostructure. This can be achieved in few-electron lateral quantum dots defined in a GaAs two-dimensional electron gas heterostructure (2DEG) where a single electron can be trapped and its spin can be measured and coherently manipulated[6]. A possible strategy to transfer spin information consists then in displacing the electron in an array of coupled quantum dots[11]. So far, only spin transfer over linear arrays of two or three dots has been demonstrated[12-15] and the complexity associated with the control of multidot systems limits the distance of propagation to a few hundred nanometres. An alternative solution consists in creating moving quantum dots using surface acoustic waves (SAWs) where the electron is trapped and propagates at the speed of sound isolated from the surrounding electrons. In previous experiments, such a strategy has been used to transfer spin information stored in an ensemble of electrons over distances larger than one hundred microns[4].

The spin transfer device consists of two lateral quantum dots, namely the source and the reception dots, linked by a 4-µm long electrostatically-depleted one-dimensional channel (see Fig. 1a). An interdigitated transducer (IDT) is placed 2 mm to the left of the gate structure. Due to the piezoelectric properties of GaAs, a 2.6326 GHz microwave excitation of the IDT resonantly generates SAWs that induce moving quantum dots propagating from the source to the reception quantum dot along the depleted channel. The principle of the experiment is as follows: one or two electrons are loaded in the source dot and different proportions of spin states are prepared by a spin relaxation process. Under the conditions of the experiment (see Supplementary Sections I and II), the relevant spin states are spin down $|\downarrow\rangle$ and spin up $|\uparrow\rangle$ in the one-electron case, parallel ($|\downarrow\downarrow\rangle$, $|\uparrow\uparrow\rangle$) and antiparallel ($|\downarrow\uparrow\rangle$, $|\uparrow\downarrow\rangle$) spin states in the two-electron case. The spin transfer procedure is then performed by generating a SAW and can be divided into three steps: (i) the electrons are first transferred from the source to the moving dots, (ii) they propagate in the moving dots over 4 µm within 1.4 ns, and (iii) they are transferred from the moving to the reception dot. Spin-selective readout is finally performed in the reception dot to infer the spin state after the spin transfer procedure.

Figure 1b presents the stability diagrams of each quantum dot. Both can be controlled down to one electron and they can easily be decoupled from their reservoirs. Such demonstrated tunability allows us to promote efficiently one (two) electrons from the loading position $L_1$ ($L_2$) to the position I where they remain isolated from the lead[16]. Subsequently, the source dot is pulsed to its transfer position T closer to the channel energy. To transfer the electrons to the reception dot, a SAW is excited by applying a microwave burst of duration $\tau_{SAW}$ to the IDT. As a result, the electrons in the source quantum dot are injected into the SAW-induced moving quantum dots and carried to the reception dot. In order to capture the electrons and keep them[9], the reception dot is tuned to the position T' where a sufficiently large dot-reservoir tunnel barrier is induced. After the end of the SAW passage, the reception dot is lowered to its isolated position I'. The successive steps of this sequence are summarized on Fig. 2a. Charge measurements before and after the transfer enable us to evaluate the transfer efficiency. The occurrence of successful transfer events for one and two electrons as a function of $\tau_{SAW}$ are presented in Fig. 2b and 2c, respectively. To limit the effect of the SAW irradiation on the spin of the trapped electrons (see Supplementary Section IV), $\tau_{SAW}$ is respectively fixed at 110 ns and 75 ns for one and two electrons.

To probe the spin dynamics of the electrons, we implemented a spin readout protocol compatible with the single-electron transfer procedure. The principle to detect electron spin states in a single quantum dot coupled to a lead is well established. It relies on the engineering of a spin-dependent tunnel process from the dot to the reservoir to convert spin into charge information. To perform single shot spin readout[17,18], we take advantage of the energy difference between the excited and the ground spin states at the measurement positions $M_1'$ ($M_2'$) for one (two) electron(s) where only electrons in the excited spin state are allowed to tunnel out of the dot (see Supplementary Section II). However, the SAW-induced electron transfer requires working with the source and the reception dots in an isolated configuration where no exchange of electrons is possible between the dot and the reservoir[9]. For this reason, we have to bring back the dot system to the measurement position at a microsecond timescale (much faster than the spin relaxation time) to infer the electron spin state after transfer.

In Fig. 2e and 2f, we evaluate the characteristics of the spin readout procedure by performing a local spin relaxation measurement in the reception dot for one and two electrons at 3 T (in the plane of the 2DEG) and at 100 mT (perpendicular to the 2DEG), respectively. By varying the waiting time before the spin readout, we expect to observe the spin relaxation process of individual electrons[18-20] and to change the population difference between the spin states. To simulate the gate voltage movement needed to readout the electron spin state after transfer, the reception dot is brought to the transfer position T' for 10 µs just before the spin readout. Clear exponential decays of the $|\downarrow\rangle$ and the parallel spin state populations are observed on a timescale of a few milliseconds, as expected for spin relaxation in single quantum dots[19, 20]. The amplitude of the exponential decay is directly connected to the measurement fidelity and is explained by the geometry of the sample and the magnetic field condition (see Supplementary Sections I and II). We therefore demonstrate a one-electron and two-electron spin-readout procedure compatible with SAW-transfer. We also performed a calibration of the spin readout protocol that will permit to evaluate the spin depolarization occurring during the electron transfer.

Finally, we combine the electron transfer and the spin readout procedures to perform a non-local spin relaxation measurement (see Fig. 3a): first, the electrons are loaded in the source dot at the position $L_{1(2)}$ for one (two) electron(s) at microsecond timescales, much faster than the spin relaxation time; second, different proportions of spin states are prepared by controlling the waiting time $t_{wait}$ before the electron transfer; third they are brought to the transfer position T; finally the electrons are conveyed to the reception dot where the spin readout is performed at the position $M_{1(2)}'$ for one (two) electron(s) just after the transfer. The probabilities to detect the electrons in the $|\downarrow\rangle$ and parallel spin states as a function of $t_{wait}$ are respectively recorded at 3 T in the plane of the 2DEG (Fig. 3b) and at 100 mT and 200 mT perpendicular to the 2DEG (Fig. 3c). We observe exponential decays of the $|\downarrow\rangle$ and parallel spin populations with relaxation times very similar to the one observed for the respective calibration experiment. This clearly demonstrates that long-range transfer of spin information using individual electrons is achieved. To quantify the efficiency of the spin transfer procedure, we compare the amplitude of the local and the non-local spin relaxation measurements. The observed amplitude reduction factors are similar in the one and two-electron cases and are independent of the magnetic field strength. We can conclude that approximately 30 % of the spin polarization initially present in the source dot has been preserved during the complete transfer procedure. In other words, assuming an equal repartition of the errors between the spin states, we obtain the error probability ε to switch between spin states during the transfer equal to 0.35 and a classical transfer fidelity 1 - ε = 0.65 (see Supplementary Section III for details).

One possibility to explain the observed loss of visibility is spin depolarization occurring during the electron propagation in the moving quantum dots. In this case, we expect the spin transfer to be characterized by individual spin-flip processes along the channel dominated by the spin-orbit interaction[21-23]. As investigated in depth in ref. 23, and contrary to what is observed in the experiment, the strength of the spin-orbit coupling in GaAs is too small to induce significant spin depolarization on a short length scale of 4 µm for external magnetic fields larger than 100 mT. Moreover, a significant difference of the contrast reduction factor is expected when spin information is encoded either in one or two electrons (see Supplementary Section III for details). These predictions are in contradiction with the experimental observations and allow us to rule out spin depolarization occurring during the electron propagation.

An alternative explanation is a spin depolarization process occurring in the static dots before and after the electron propagation. During the transfer between the static and the moving dots, the electrons are indeed experiencing an important perturbation due to the SAW excitation. Because of the presence of a double dot potential close to the transfer position[16] (see Supplementary Sections I and II), multiple passages through spin level anti-crossings induced by either spin-orbit or hyperfine coupling[25] likely occur during the SAW excitation and result in spin mixing. A series of experimental results supporting this scenario are presented in the Supplementary Section IV.

In conclusion, we have demonstrated that spin information can be transferred between distant quantum dots separated by 4 µm using either one or two electrons. Spin polarization and spin readout of the electron spin state are performed in static dots whereas the transfer is mediated by moving quantum dots generated in a long depleted channel with SAWs. The classical fidelity of the whole spin transfer procedure reaches 65 % and is believed to be limited by the depolarization in the static dots prior to and after the electron displacement. In comparison with previous demonstration of electron spin transfer, the time of the electron transfer is much shorter than the spin coherence time and therefore our technique could, according to theoretical predictions[23], be used to demonstrate coherent spin transfer. It will require combining already demonstrated nanosecond control of the electron transfer[9] and nanosecond spin manipulation of the electron spin states close to the transfer position[16]. It would potentially permit to realize on-chip quantum teleportation protocols for electron spins in quantum dots over distances larger than one hundred microns[4].

**References**

1. Dlubak, B. *et al.* Highly efficient spin transport in epitaxial graphene on SiC. *Nature Physics,* **8**, 557-561 (2012).

2. Lou, X. *et al.* Electrical detection of spin transport in lateral ferromagnet-semiconductor devices. *Nature Physics,* **3**, 197-202 (2007).

3. Kato, Y., Myers, R., Gossard, A. & Awschalom, D. Coherent spin manipulation without magnetic fields in strained semiconductors. *Nature,* **427**, 50-53 (2004).

4. Stotz, J. A. H., Hey, R., Santos, P. V. & Ploog, K. H. Coherent spin transport through dynamic quantum dots. *Nature Materials,* **4**, 585-588 (2005).

5. Sanada, H. *et al.* Manipulation of mobile spin coherence using magnetic-field-free electron spin resonance. *Nature Physics,* **9**, 280-283 (2013).

6. Hanson, R., Petta, J. R., Tarucha, S. & Vandersypen, L. M. K. Spins in few-electron quantum dots. *Reviews of Modern Physics,* **79**, 1217-1265 (2007).

7. Shulman, M. D. *et al.* Demonstration of entanglement of electrostatically coupled singlet-triplet qubits. *Science,* **336**, 202-205 (2012).

8. Veldhorst, M. *et al.* An addressable quantum dot qubit with fault-tolerant control-fidelity. *Nature Nanotechnology,* **9**, 981-985 (2014).

9. Hermelin, S. *et al.* Electrons surfing on a sound wave as a platform for quantum optics with flying electrons. *Nature,* **477**, 435-438 (2011).

10. McNeil, R. *et al.* On-demand single-electron transfer between distant quantum dots. *Nature,* **477**, 439-442 (2011).

11. Thalineau, R. *et al.* A few-electron quadruple quantum dot in a closed loop. *Applied Physics Letters,* **101**, 103102 (2012).

12. Petta, J. R. *et al.* Coherent manipulation of coupled electron spins in semiconductor quantum dots. *Science,* **209**, 2180-2184 (2005).

13. Medford, J. *et al.* Quantum-dot-based resonant exchange qubit. *Phys. Rev. Lett.,* **111**, 050501 (2013).

14. Sanchez, R. *et al.* Long-range spin transfer in triple quantum dots. *Phys. Rev. Lett.* **112**, 176803 (2014).

15. Baart, T. A. *et al*. A single spin CCD, *Nature Nanotechnology* (2016)

16. Bertrand, B. *et al*. Quantum manipulation of two-electron spin states in metastable double quantum dots. *Phys. Rev. Lett.* **115**, 096801 (2015).

17. Meunier, T. *et al.* High fidelity measurement of singlet-triplet state in a quantum dot. *Physica status solidi (b),* **243**, 3855-3858 (2006).

18. Elzerman, J. M. *et al.* Single-shot read-out of an individual electron spin in a quantum dot. *Nature,* **430**, 431-435 (2004).

19. Amasha, S. *et al.* Electrical control of spin relaxation in a quantum dot. *Phys. Rev. Lett.,* **100**, 046803 (2008).

20. Meunier, T. *et al.* Experimental signature of phonon-mediated spin relaxation in a two-electron quantum dot. *Phys. Rev. Lett.,* **98**, 126601 (2007).

21. Zutic, I., Fabian, J., & Das Sarma S., *Rev. Mod. Phys.* 76, 323 (2004)

22. Dugaev, V. K., Sherman, E. Y., Ivanov, V. I., & Barnas, J., *Phys. Rev. B* 80, 081301 (2009)

23. Huang, P. & Hu, X. Spin qubit relaxation in a moving quantum dot. *Phys. Rev. B* **88**, 075301 (2013).

24. Meier, L. *et al.* Measurement of Rashba and Dresselhaus spin-orbit magnetic fields. *Nature Physics* 3, 650 - 654 (2007).

25. Stepanenko, D., Rudner, M., Halperin, B. I. & Loss, D. Singlet-triplet splitting in double quantum dots due to spin-orbit and hyperfine interactions. *Phys. Rev. B* 85, 075416 (2012).

**Acknowledgements**

We acknowledge technical support from the technological centres of the Institut Néel. Sh.T. acknowledges financial support from the Marie Sklodowska-Curie grant agreement No 654603. M.Y. and Se. T. acknowledge financial support by JSPS (Grant No. 26247050, No. 25610070 and No. 26220710). Se. T. acknowledges financial support by MEXT KAKENHI "Quantum Cybernetics," MEXT project for Developing Innovation Systems, and JST Strategic International Cooperative. A.L. and A.D.W. acknowledge gratefully the support of the BMBF Q.com-H 16KIS0109, Mercur Pr-2013-0001 and the DFH/UFA CDFA-05-06. B. B. and T. M. acknowledge financial support from ERC "QSPINMOTION" and the Fondation Nanosciences.

**Contributions**

B.B. performed the experiments. B.B. and T.M. interpreted the data. B.B., C. B. and T.M. wrote the manuscript. Sh. T. designed and fabricated the sample with M.Y. and Se.T. S.H. contributed to the experimental setup. A. L. and A.D.W. provided the high mobility heterostructures. All authors discussed the results extensively as well as the manuscript.

**Competing financial interests**

The authors declare no competing financial interests.

**Corresponding authors**

Correspondence and requests for materials should be addressed to T.M. (tristan.meunier@neel.cnrs.fr).

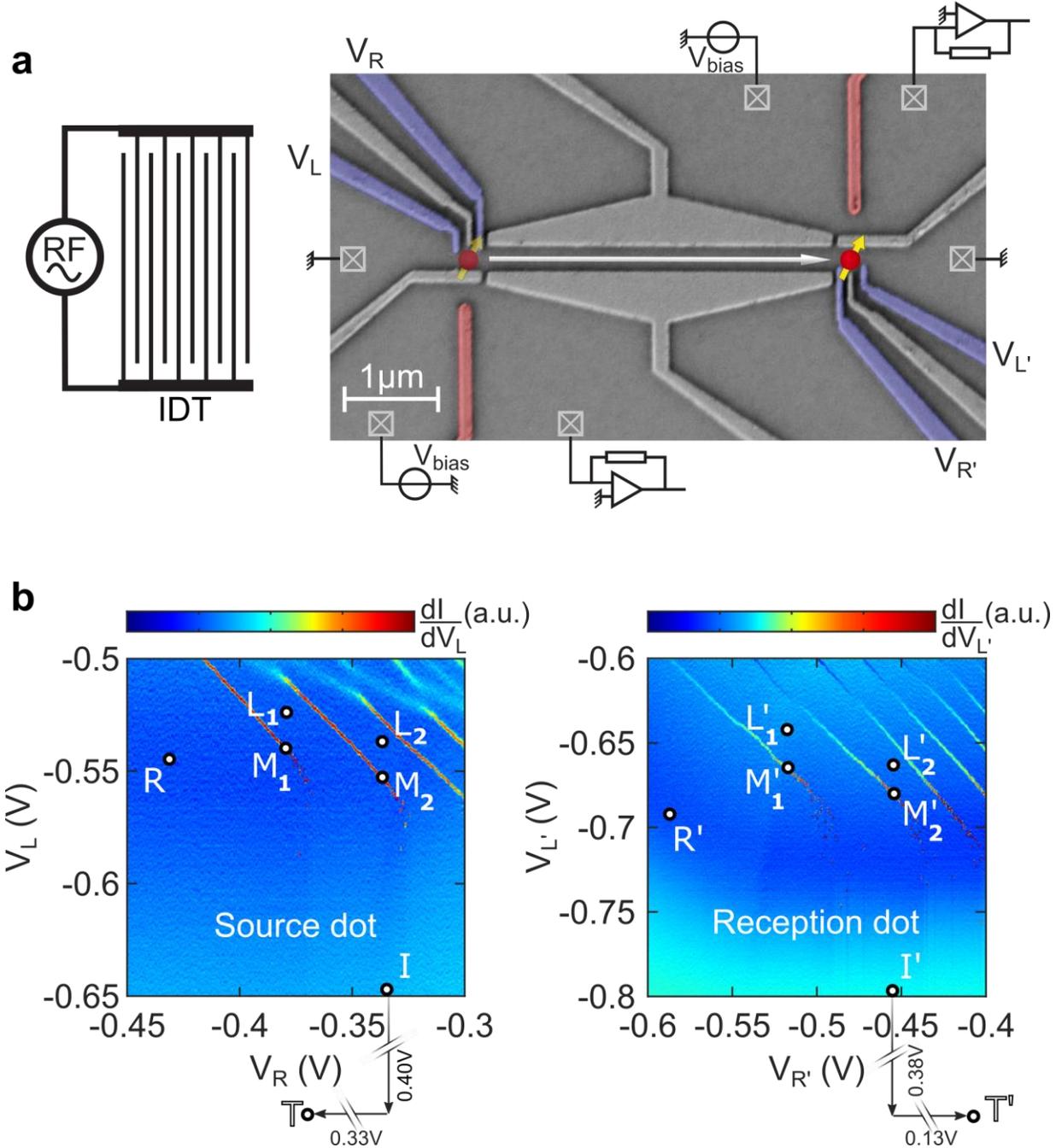

**Figure 1 Experimental set-up and characterizations of the quantum dots. a,** Scanning electron microscope image of the spin transfer device. It is fabricated using a GaAs/AlGaAs heterostructure grown by molecular beam epitaxy with a two-dimensional electron gas (2DEG) 100 nm below the surface. It is formed by local depletion of the 2DEG by means of metal Schottky gates deposited on the surface of the sample. The dot properties can be changed by varying the voltages applied to the blue surface gates at microsecond timescales (see Supplementary Section I for details). The charge state of each dot can be monitored using an on-chip electrometer (a quantum point contact) defined with the red gate with a ~1 kHz detection bandwidth imposed by room-temperature electronics. **b,** Left (Right): Stability diagram of the source (reception) dot. One and two electrons can be loaded in 10 µs either in the source (reception) dot respectively at loading positions $L_1$ ($L_1'$) and $L_2$ ($L_2'$). The spin readout of the one and two electron spin states are performed in the reception dot at measurement positions $M_1'$ and $M_2'$. The electrons are transferred efficiently from transfer positions T of the source dot to T' of the reception dot and the charge measurements before and after transfer are respectively performed in the isolated position I of the source dot and I' of the reception dot.

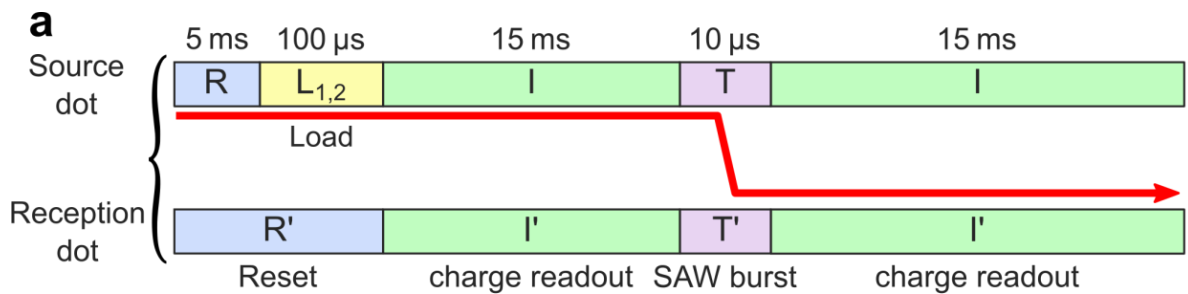

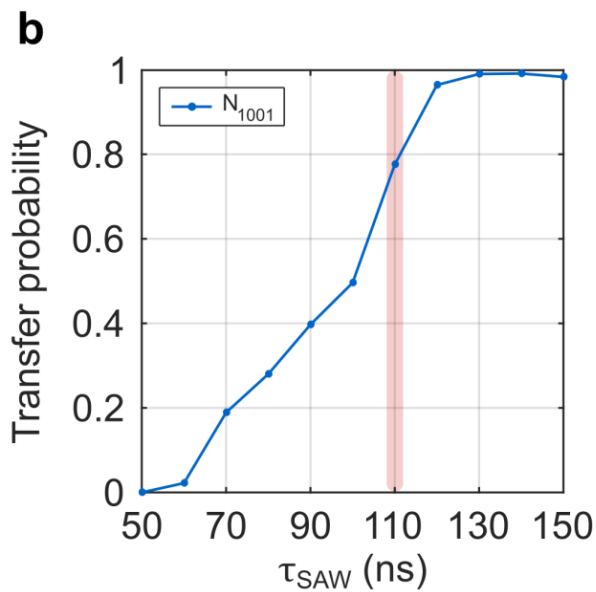
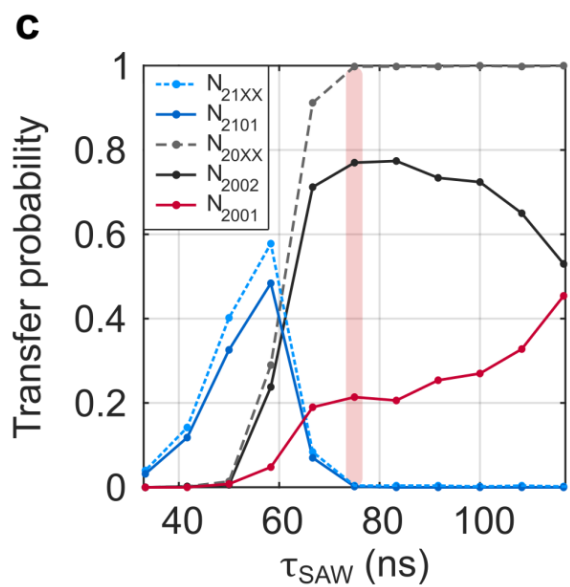

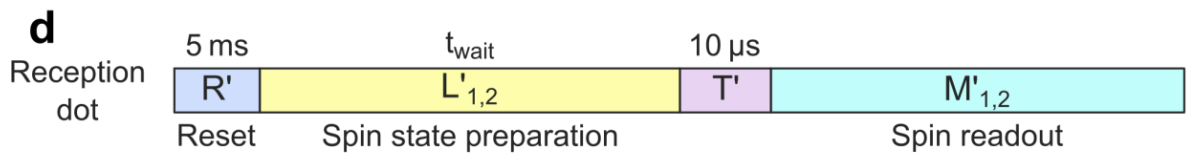

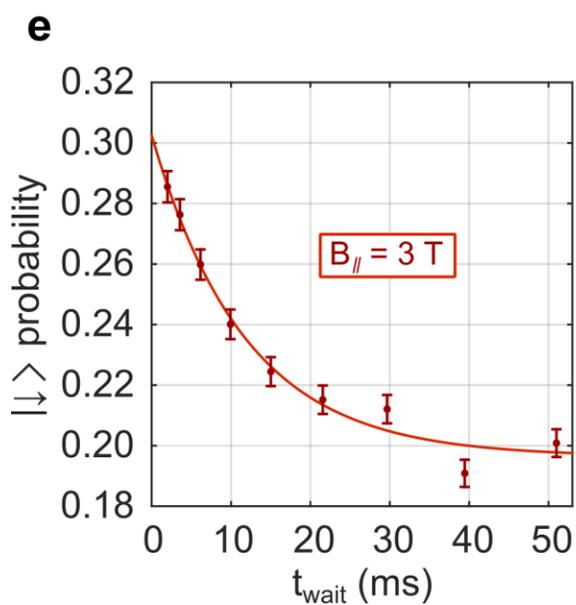
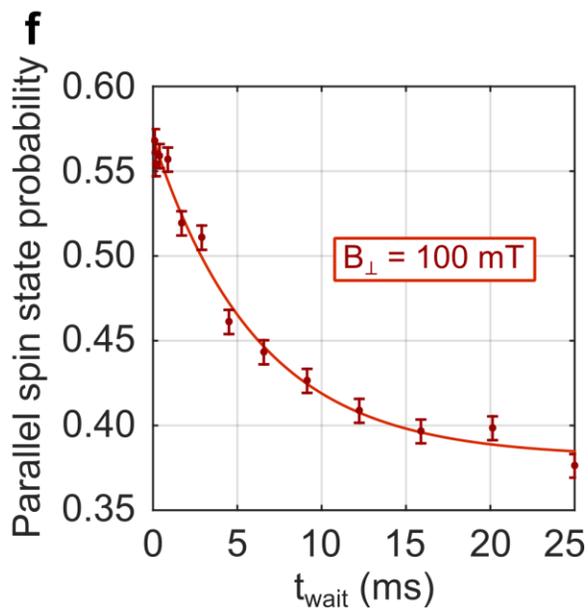

**Figure 2 Local spin relaxation measurements in the reception dot and electron transfer between the source and the reception dots. a,** Schematics of the successive positions of the source and reception quantum dots during the transfer sequence used in Fig. 2b and 2c. **b,** Single electron transfer probability as a function of the SAW burst duration $\tau_{SAW}$. **c,** Two-electron transfer probability as a function of $\tau_{SAW}$. $N_{\alpha\beta\gamma\delta}$ refers to the counts for the source (reception) dot loaded with α (γ) electrons, ending up with β (δ) electrons after the SAW burst. The total number of recorded traces is 1000 in 2b and 500 in 2c. In both situations, a sufficiently large $\tau_{SAW}$ is required to emit the electrons. In the two-electron case, only the first electron is transferred at short $\tau_{SAW}$, whereas both electrons are transferred at longer $\tau_{SAW}$. It demonstrates that the electrons are propagating in different moving quantum dots. The $\tau_{SAW}$ values used for the non-local spin relaxation measurement are highlighted by the red vertical bars in Fig. 2b and 2c. **d,** Schematics of the successive positions of the reception dot for the local spin relaxation measurements shown in Fig. 2e and 2f. In both experiments, a 10 µs pulse to position T' is performed just before the spin readout (see Supplementary Section II for details). **e,** $|\downarrow\rangle$ probability as a function of the waiting time $t_{wait}$ between the electron loading and the spin readout in the reception dot. An in-plane magnetic field of magnitude B = 3 T is applied. The curve corresponds to a fit with an exponential decay A×exp(-t/$T_1$) + B. We obtain an amplitude A = 0.108 ± 0.014 and a spin relaxation time $T_1$ = 11.8 ms ± 5.7 ms. **f,** Parallel spin state probability as a function of $t_{wait}$ at 100 mT perpendicular magnetic field. The exponential decay fit gives A = 0.188 ± 0.016 and $T_1$ = 6.2 ms ± 1.6 ms. Due to the similarity between the source and reception dots, similar spin dynamics have been observed in the source dot (see Supplementary Section II).

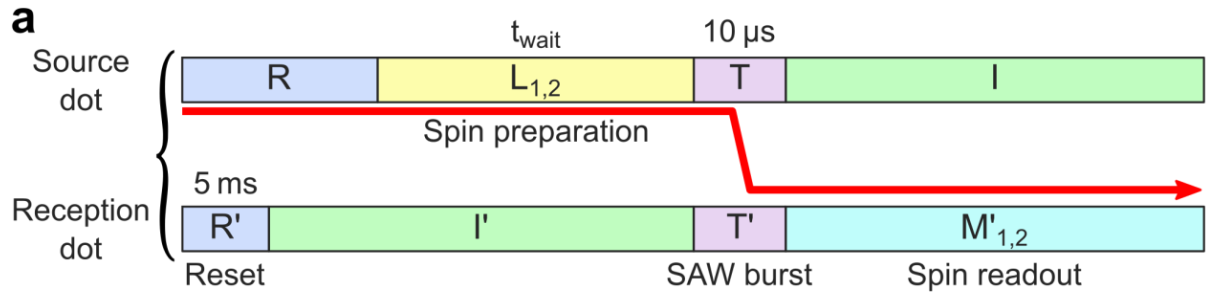

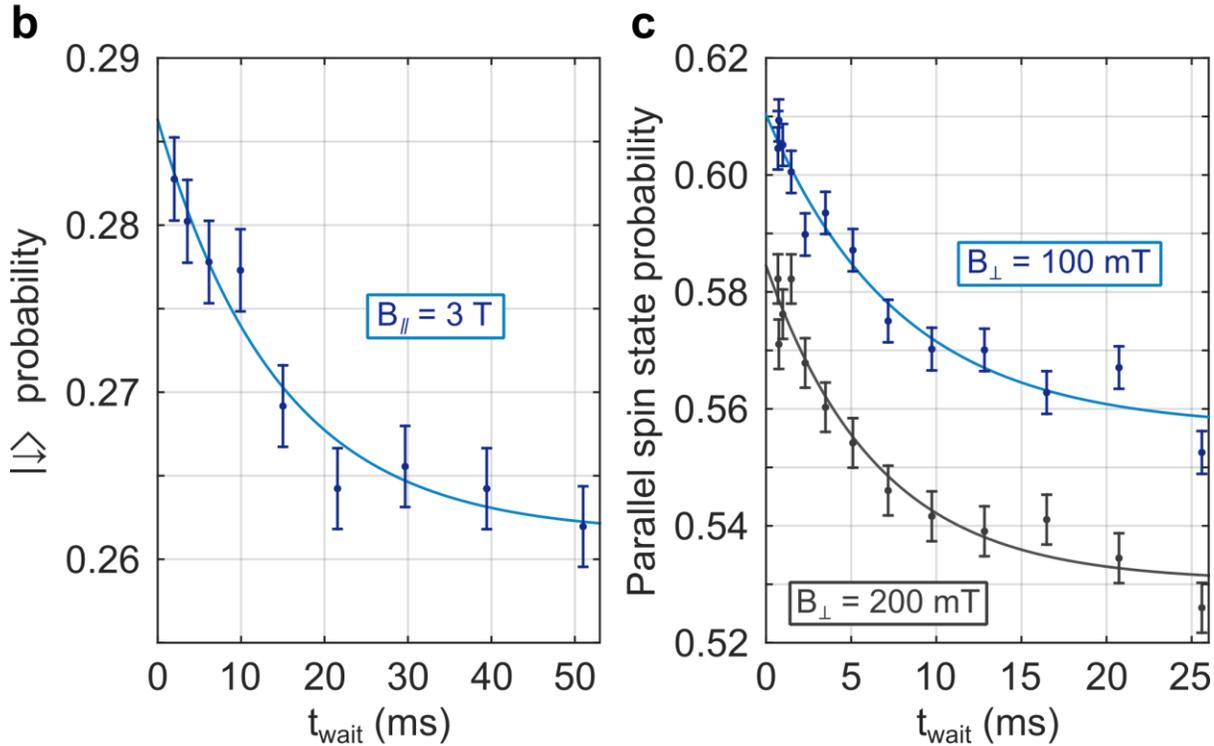

**Figure 3 Non local spin relaxation measurements. a,** Schematics of the successive positions of the source and reception quantum dots for the non-local spin relaxation measurements shown in Fig. 3b and 3c. **b,** $|\downarrow\rangle$ probability after the single electron transfer ($\tau_{SAW}$ = 110 ns) as a function of $t_{wait}$ at B = 3 T (in-plane). The exponential fit gives a signal amplitude of A = 0.025 ± 0.005 and a spin relaxation $T_1$ = 14.5 ms ± 10 ms. **c,** Parallel spin state probability after the two-electron transfer ($\tau_{SAW}$ = 70 ns) as a function of $t_{wait}$ with a perpendicular magnetic field of 100 mT (blue) and 200 mT (black). The blue curve has been shifted vertically by + 0.01 for clarity. The resulting fit

parameters are A = 0.054 ± 0.01 and $T_1$ = 7.7 ms ± 4.2 ms for the blue curve and A = 0.054 ± 0.009 and $T_1$ = 6.5 ms ± 3.6 ms for the black curve.

**METHODS**

The device is defined by Schottky gates in an n-Al$_{0.3}$Ga$_{0.7}$As/GaAs 2DEG-based heterostructure (the properties of the non-illuminated 2DEG are as follows: $\mu \approx 10^6$ cm$^2$ V$^{-1}$ s$^{-1}$, $n_s \approx 1.4 \times 10^{11}$ cm$^{-2}$, depth 90 nm) with standard split-gate techniques. It is anchored to a cold finger mechanically screwed to the mixing chamber of a dilution fridge with a base temperature of 50 mK. It is placed at the centre of the magnetic field produced by a 2D magnet. The magnet allows to produce magnetic fields perpendicular or in the plane of the 2DEG. The charge configuration of both dots is measured by means of the conductance of both QPCs by biasing it with a direct-current voltage of 300 µV; the current is measured with a current-to-voltage converter with a bandwidth of 1.4 kHz. The voltage on each gate can be varied on a timescale down to microseconds. The IDT, which is placed about 2 mm to the left of the sample, is made of 70 pairs of lines 70 µm-long and 250 nm-wide with 1 µm spacing. The fingers of the IDT are oriented perpendicular to the direction of the 1D channel defined along the crystal axis [110] of the GaAs wafer. The IDT is then generating SAWs propagating along the channel direction. The resonance frequency is 2.6326 GHz with a bandwidth of about 2.4 MHz[26].

26. Bertrand B. *et al.*, Injection of a single electron from a static to a moving quantum dot, arXiv:1601.02485, Nanotechnology, in press.

Supplementary information for

# Fast spin information transfer between distant quantum dots using individual electrons

B. Bertrand, S. Hermelin, S. Takada, M. Yamamoto, S. Tarucha, A. Ludwig, A. D. Wieck, C. Bäuerle, T. Meunier*

# Content

# I. Emergence of a double dot behavior in the isolated regime

As mentioned in the main manuscript, different positions are used during the transfer sequence, with significant changes in terms of sample tuning. For the electron loading procedure and the spin measurement, exchange of electrons with the reservoirs is required. The dot-reservoir couplings are usually tuned between 1 kHz and a few 10 kHz. On the contrary, in the position used for the electron transfer, exchange of electrons with the reservoirs must be suppressed. For this purpose, the voltages applied on gates $V_L$ and $V_{L'}$ are set much more negative to suppress the dot-reservoir couplings (<<1 kHz). In parallel, this raises the chemical potential of the dots closer to that of the channel, and therefore facilitates the electron transfer.

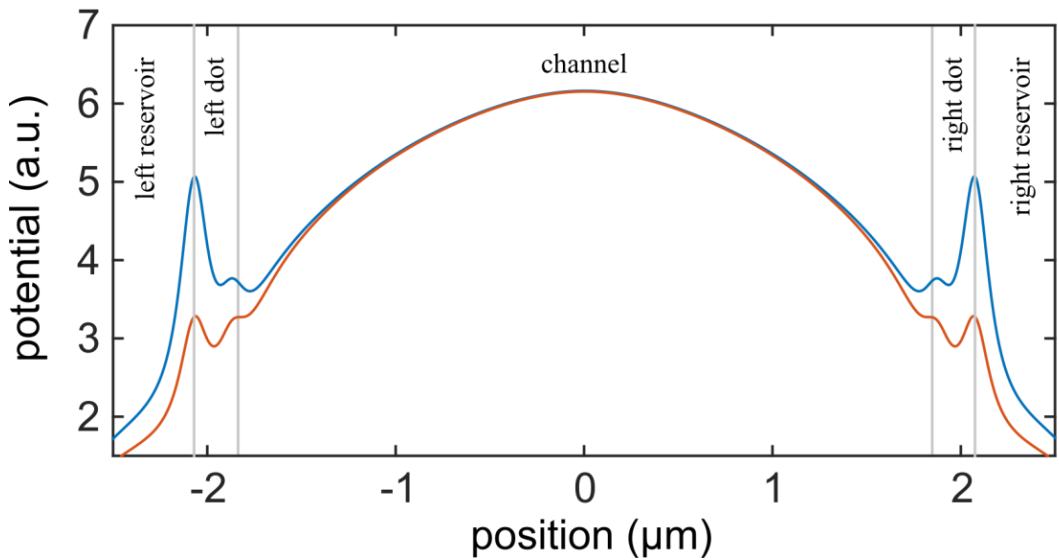

**Supplementary Fig. 1. Numerical simulations of the potential.** The red (blue) curve corresponds to a slice of the potential calculated for gate voltages for the loading (transfer) position. While there is hardly a double dot behavior at the loading position, it clearly emerges in the transfer position.

Numerical simulations of the potential profile following the reference 27 have been realized for both the coupled and the isolated regimes, and the results are shown on Supplementary Fig. 1. Because of the sample geometry, a local minimum can show up between the gate $V_R$ (or symmetrically $V_{R'}$) and the channel gates. It remains out of reach while in the coupled regime (red curve), but becomes accessible if the system is pulsed to the isolated regime (blue curve). A precise study of these isolated double dots has been done[16] and interesting results in terms of coherent spin manipulations have been obtained. On the one hand, this double dot behavior could be useful to investigate the transfer of the two-electron "$|S\rangle - |T_0\rangle$" spin qubit. On the other hand, it is the main limitation for the spin transfer fidelity in our experiment and we will detail this point in the following sections.

## II. Spin relaxation and relevant spin states in the experiment

First, we explain in this section the low contrast obtained for the two-electron spin relaxation in the reception dot (see Fig. 2f of the main text). Energy-selective readout of two-electron spin states allows one to distinguish between the singlet ground state $|S\rangle = \frac{|\uparrow\downarrow\rangle - |\downarrow\uparrow\rangle}{\sqrt{2}}$ and the three excited triplet states $|T_+\rangle = |\uparrow\uparrow\rangle$, $|T_0\rangle = \frac{|\uparrow\downarrow\rangle + |\downarrow\uparrow\rangle}{\sqrt{2}}$, and $|T_-\rangle = |\downarrow\downarrow\rangle$.

To obtain high contrast singlet-triplet relaxation curves[17], the experimental protocol is as follows: two electrons are loaded in the reception dot at the position L$_2$' and the electrons are mostly prepared in the triplet states due to the tunnel rate difference between the singlet and triplet states[20]; after a waiting time t$_{wait}$ at this position, the system is then brought at a microsecond timescale to position M$_2$' to perform single-shot energy selective readout (see Supplementary Fig. 2a). Following this protocol, we measure the relaxation signal represented in green on Supplementary Fig. 2b. It shows a higher contrast than the one reported in Fig. 2f, approaching 70 % as can be expected from the measurement bandwidth of our set-up. This curve has been obtained with a 100 mT perpendicular magnetic field applied, but a similar contrast is observed when switching the magnetic field to 0 mT. In this way, we probe the relaxation process from the triplet states ($|T_+\rangle$, $|T_0\rangle$, $|T_-\rangle$) to the singlet state $|S\rangle$.

In the experiment presented in Fig. 2f of the main manuscript, the experimental protocol is slightly different: two electrons are loaded in the reception dot at the position L$_2$' and after the waiting time t$_{wait}$, a 10 μs pulse to the transfer position T' is added just before going to the measurement position M$_2$'. This pulse to the transfer position is added for a better comparison with the non-local spin relaxation measurement protocol. In this way, we calibrate the maximal expected spin relaxation signal after electron transfer. The results obtained with this second protocol are shown as the red and purple curves on Supplementary Fig. 2b, corresponding respectively to 100 mT and 0 mT of applied magnetic field. At B = 100 mT, the contrast of the spin relaxation curve drops to 19 %, and no contrast at all is observed at B = 0 mT. These observations are consistent with the double dot behavior presented in the previous section. The electron pair is separated in two weakly coupled quantum dots during the pulsing to the transfer position T', allowing the coupling to the nuclear spins of the heterostructure to induce mixing between the two-electron spin states[28]. At zero magnetic field, all the four two-electron spin states are degenerate and mix, whereas only $|T_0\rangle$ and $|S\rangle$ mix at 100 mT. As a result, after pulsing the dot to its transfer position T' and with a magnetic field larger than 100 mT, we are only able to probe the proportion of parallel ($|\uparrow\uparrow\rangle$ and $|\downarrow\downarrow\rangle$) and antiparallel ($|\uparrow\downarrow\rangle$ and $|\downarrow\uparrow\rangle$) spin states. Similar spin dynamics is expected and observed in the source dot due to the very similar geometry (see Supplementary Fig. 2b).

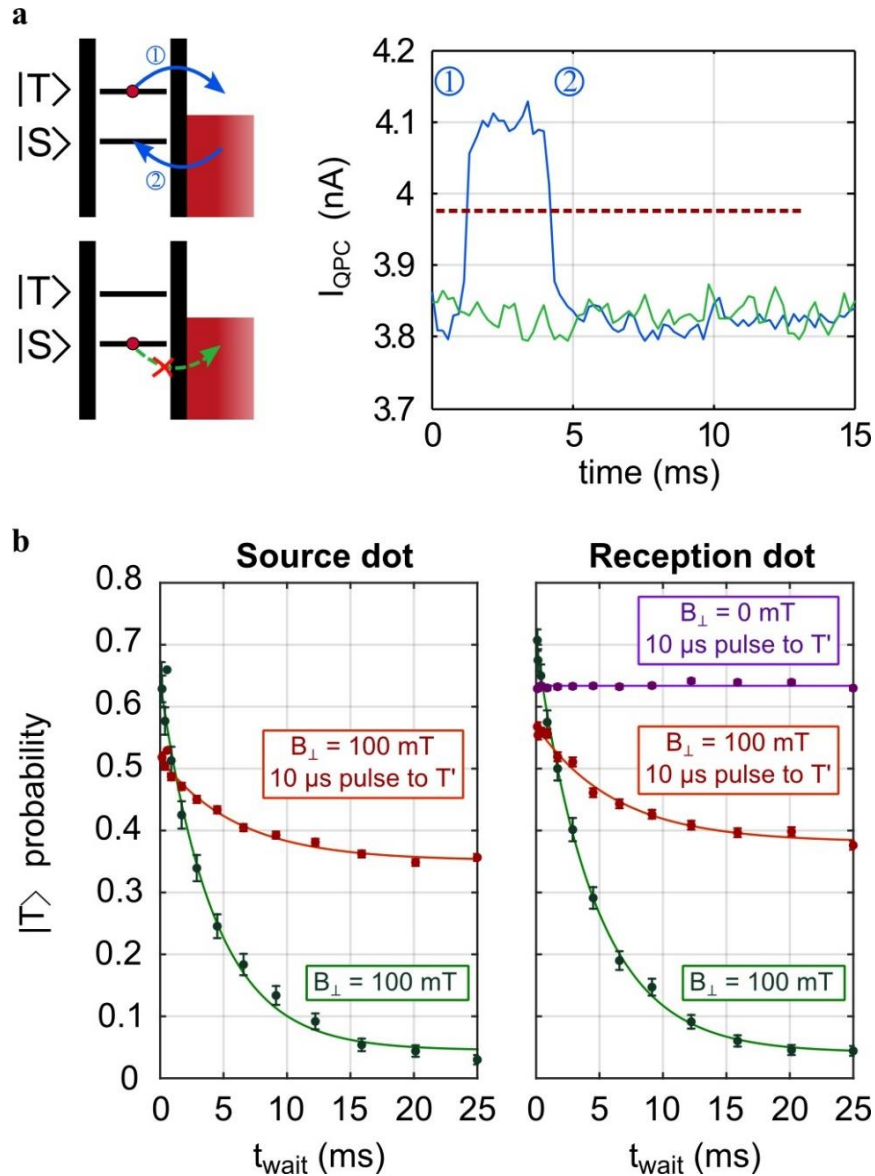

**Supplementary Fig. 2. Local two-electron spin relaxation measurement in the source and reception dot. a,** Sketch of the energy selective spin readout procedure used for the spin relaxation measurements, and example of the typical time traces observed for excited (blue) or ground (green) spin states. **b,** Green: Triplet probability as a function of the waiting time $t_{wait}$ at position $L_2'$ between the electron loading and the spin readout. A 100 mT perpendicular magnetic field is applied. The same result is observed at zero magnetic field. Red and purple: similar measurement with the addition of a 10 µs pulse to the position T' just before the spin readout, with 100 mT perpendicular magnetic field applied (red) and at zero magnetic field (purple). Similar spin dynamics is expected and observed in the source dot due to the very similar geometry.

It is worth noting that for the single electron case, this additional pulse to the transfer position caused no drastic change in the spin relaxation curve contrast as expected, and the contrast in this case is directly explained by the magnetic field and the bandwidth of the electronics.

# III. Evaluation of the fidelity of the transfer procedure

We demonstrated that one and two–electron spin states can be transferred between distant dots. In this section, we give a detailed derivation of the extracted classical fidelities $f_1$ and $f_2$ of the one and two-electron spin transfer procedures and their relations to the individual spin-flip probability $p$, assuming a spin depolarization process during the displacement of the electron. In both cases, we assume $p$ independent of the initial spin state.

The amplitude reduction factor $c_i$ is obtained by dividing the extracted amplitude of the non-local spin measurement (Fig. 2) with the one of the local spin measurement (Fig. 3). We obtained $c_1 = 0.23 \pm 0.07$ and $c_2 = 0.29 \pm 0.07$ respectively for the one and two-electron cases.

The error probabilities $\varepsilon_1$ and $\varepsilon_2$ of the transfer procedure are respectively defined as the probability to switch between $|\uparrow\rangle$ and $|\downarrow\rangle$ in the one-electron case and between parallel ($|\uparrow\uparrow\rangle$ and $|\downarrow\downarrow\rangle$) and antiparallel spin states ($|\uparrow\downarrow\rangle$ and $|\downarrow\uparrow\rangle$) in the two-electron case (see Supplementary Fig. 3 for details). Such errors result in a reduction of the spin relaxation curve amplitude. The amplitude reduction factor $c_i$ is therefore linked to $\varepsilon_i$ via the relation $c_i = 1-2\varepsilon_i$ and we then obtain from the data $\varepsilon_1 = 0.38 \pm 0.04$ and $\varepsilon_2 = 0.35 \pm 0.04$. To evaluate the fidelities for each transferred spin state, we use the definition of fidelity in quantum information theory[29]. The fidelity between two density matrices σ and ρ is defined as $f(\sigma,\rho) = \left[Tr\left(\sqrt{\sqrt{\sigma}\rho\sqrt{\sigma}}\right)\right]^2$. Therefore the fidelities for $|\uparrow\rangle$ and $|\downarrow\rangle$ are equal to $f_1 = 1 - \varepsilon_1 = 0.62 \pm 0.04$; the fidelities for parallel and antiparallel two-electron spin states are equal to $f_2 = 1 - \varepsilon_2 = 0.65 \pm 0.04$.

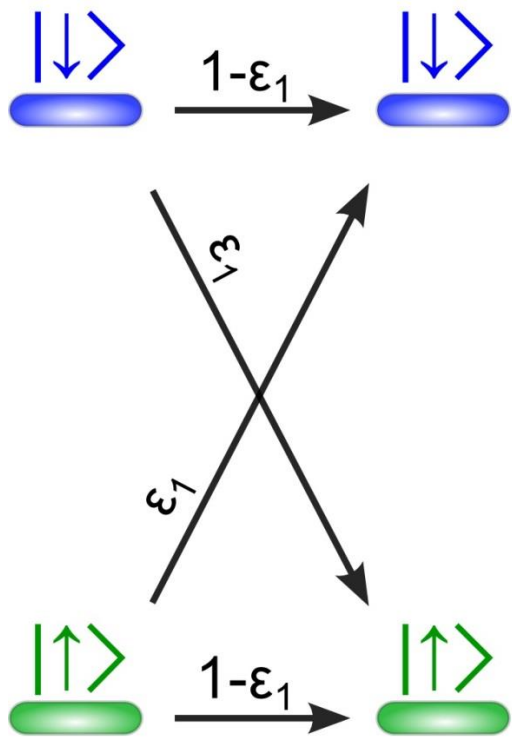 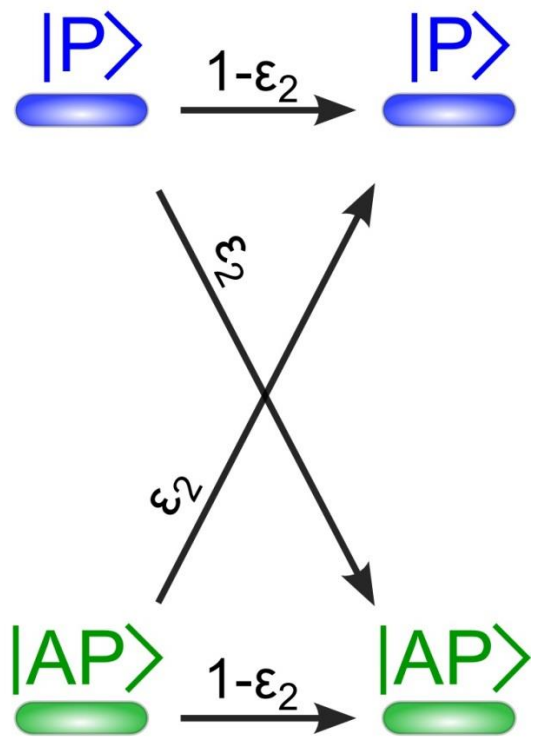

**Supplementary Fig. 3. Fidelity parameters of the spin transfer procedure for one and two electrons.** The individual spin-flip probability *p* is the probability for an individual electron to flip its spin states during the transfer procedure. In the one electron case, $\varepsilon_1 = p$. For two electrons, switching between $|AP\rangle$ (antiparallel) and $|P\rangle$ (parallel) spin states requires that the first electron flips its spin whereas the second remains unchanged, or *vice versa*. Therefore, $\varepsilon_2 = 2p(1-p)$.

We consider now the case of spin-flips occurring during the electron propagation. We assume an individual spin-flip process even for the two-electron spin states, since the electrons are propagating in different moving quantum dots. For magnetic fields larger than 100 mT, ref. 23 demonstrates that the spin-flip probability is independent of the external magnetic field. We can then relate the error probabilities $\varepsilon_1$ and $\varepsilon_2$ to the same individual spin–flip probability p as described in the Supplementary Fig. 3. Whereas *p* is directly related to the error probability $\varepsilon_1$ in the one-electron case, the relation is more complex in the two-electron case and the error probability $\varepsilon_2$ is then $2p(1-p)$. Taking p extracted from the one-electron experiment, a contrast reduction factor $c_2 = 0.08$ is expected for the two-electron experiment. This value is much lower than what is observed in the experiment. It constitutes an indication that the spin-flip process does not occur during the electron propagation.

## IV. Evidence of spin mixing induced by SAWs in an isolated double dot with one electron

In the main manuscript, we describe an alternative explanation to the loss of spin information during the transfer. It consists of a spin depolarization process occurring before and after the transfer in the source and/or reception dots. The SAW excitation indeed induces an important perturbation when electrons are trapped in the static dots that could lead to spin depolarization. In this section, we report on experiments where either one or two electrons are kept within a given static dot during the SAW excitation, and their spin states are measured after SAW irradiation. The experimental results support the depolarization mechanism pointed out in the manuscript.

We first performed an experiment where a single electron is kept in the reception dot and experiences a SAW irradiation of varying duration $\tau_{SAW}$. The electron is first loaded in the reception dot at the position $L_1'$. After a waiting time $t_{wait}$ in this position, a 10 μs pulse brings the electron to the transfer position T'. The SAW irradiation is applied during this pulse. Finally the system is brought to position $M_1'$ to perform single shot spin readout. In Supplementary Fig. 4a we present the detected probability of $|\downarrow\rangle$ as a function of the SAW excitation duration $\tau_{SAW}$.

It has been measured for two different values of $t_{wait}$, either short (blue curve) or long (green curve) relative to the single electron spin relaxation time. The spin states initialized in the dot for short or long $t_{wait}$ are expected to be an almost equal mixture of the $|\uparrow\rangle$ and $|\downarrow\rangle$ states or a pure $|\uparrow\rangle$ state, respectively. In that way, the amplitude of the spin relaxation measurement can be directly inferred from the difference between both curves.

Without SAW excitation ($\tau_{SAW}$ = 0 ns), we recover roughly the 10 % amplitude of the spin relaxation curve presented in the Fig. 2e of the main manuscript. As $\tau_{SAW}$ is increased, the signal difference between short and long $t_{wait}$ is progressively reduced. We interpret it as the result of the SAW excitation in the geometrical configuration of the sample. As demonstrated in the previous section as well as in previously reported measurements[16], an isolated double dot potential is present close to the transfer position, with the two dots aligned along the SAW propagation direction. In this situation, the SAW excitation mainly affects the relative detuning between the two dots that can result in the passage through anticrossings between excited and ground spin states of the two dots induced by either spin-orbit interaction or hyperfine coupling to the nuclei of the heterostructure (see Supplementary Fig. 4b). During the SAW excitation, the electron is experiencing several hundreds of Landau-Zener transitions through these level anticrossings, leading to extra mixing between excited and ground spin states. This would result in a reduction of the spin relaxation curve amplitude as observed in Fig. 3.

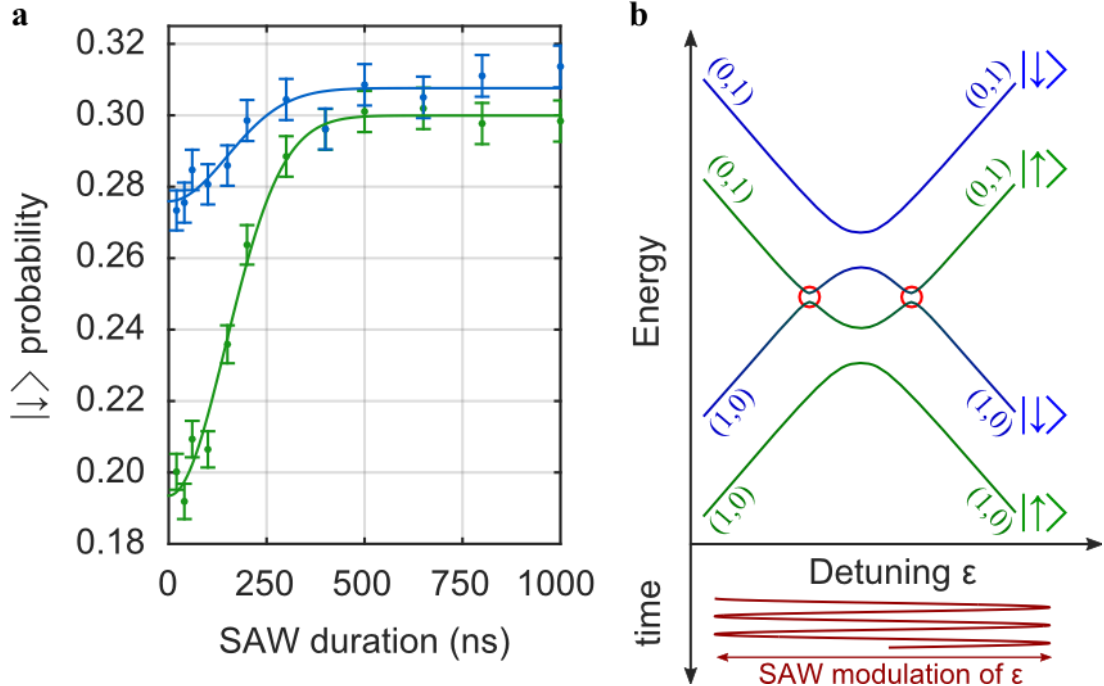

**Supplementary Fig. 4. SAW effect on the single electron spin state in a quantum dot. a**, $|\downarrow\rangle$ probability as a function of $\tau_{SAW}$ at the transfer position T' of the reception dot at 3 T in the plane of the 2DEG. A single electron spin is loaded at position $L_1'$ and is promoted to the transfer position T' in the reception dot. Its polarization is changed by controlling the waiting time between the loading and the spin readout. The electron is experiencing SAW excitation just before spin readout. Due to the presence of the large barrier to the reservoir, the electron remains in the reception dot all along the SAW excitation. The blue (green) curve corresponds to the electron brought to the transfer position T' right after (50 ms after) the loading. After 250 ns SAW excitation, the spin states are completely mixed. **b**, Energy levels of the one electron in the isolated double dot and schematics of the SAW excitation. The SAW duration dependence of the blue curve can be explained by two effects: either the unavoidable relaxation during the gate movement to go to the measurement position (about 1 ms) or a spin-dependent tunneling in the dot which slightly favors $|\uparrow\rangle$ electron to enter in the dot[17].

We performed a second experiment where the SAW influence is investigated on two-electron spin states in the source quantum dot. The idea is to demonstrate that SAW-induced passages through spin level anticrossings are indeed harmful for the spin state of the system. Following the work presented in reference 16, a precise characterization of the isolated double quantum dot that emerges close to the transfer position has been done. It allows us to determine precisely the location in the gate voltage space of the $|S\rangle - |T_+\rangle$ anticrossing. To study the influence of the SAW, the following experimental protocol is used: first, a singlet ground state is initialized by loading two electrons in the position $L_2$ and by waiting a time $t_{wait} = 25$ ms, much longer than the spin relaxation time. Next, the system is pulsed to the isolated region, away from the $|S\rangle - |T_+\rangle$ anticrossing by a value $\Delta V_{S-T+}$ in gate voltage (controlled by the gate $V_L$). The SAW excitation is then applied for 100 ns at this position. Finally, the system is set back to the measurement position to readout the spin states after the SAW excitation. This experiment has been realized for different SAW amplitudes and at positions with different values of $\Delta V_{S-T+}$. The measured spin state probability as a function of these two parameters is presented in Supplementary Fig. 5.

For a low SAW excitation amplitude and large enough distance $\Delta V_{S-T+}$ from the $|S\rangle - |T_+\rangle$ anticrossing, the spin state is not affected and the measured triplet probability is close to 0 (bottom right half of the diagram in dark blue). As the system is pulsed closer to the $|S\rangle - |T_+\rangle$ anticrossing (by increasing the SAW amplitude), a higher triplet probability is measured indicating that spin mixing occurs (top left half of the diagram). From these results, we attribute the spin mixing to the fact that the potential modulation induced by the SAW excitation becomes sufficient to induce passages through the $|S\rangle - |T_+\rangle$ anticrossing. This observation suggests that the main source of spin mixing occurring here is indeed passages through the spin level anticrossings.

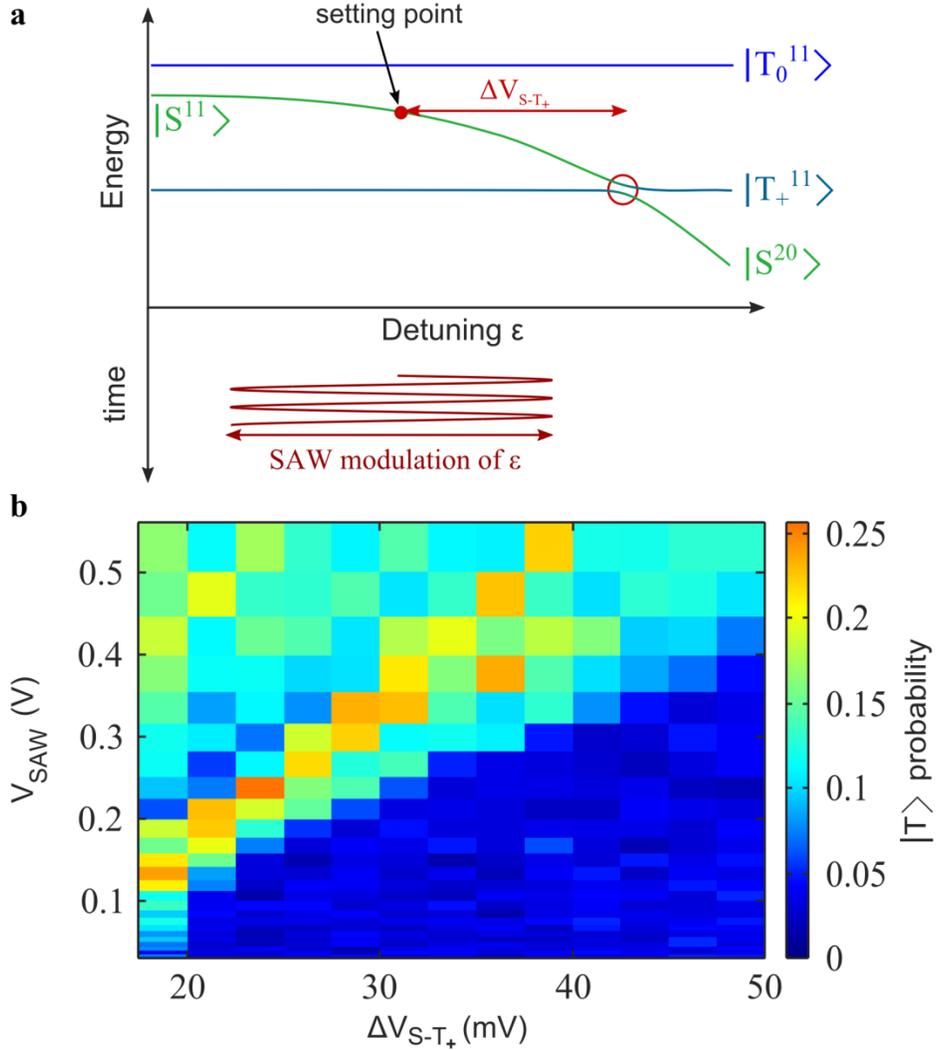

**Supplementary Fig. 5. Evidence of spin mixing caused by SAW-induced passages through a spin level anticrossing for a two-electron spin state. a,** Energy levels of two electrons in the isolated double dot around the studied $|S\rangle - |T_+\rangle$ anticrossing (red circle), and schematics of the SAW excitation. A singlet state is initially prepared in a position detuned by $\Delta V_{S-T+}$ away from the anticrossing (setting point). **b,** Triplet probability after SAW irradiation as a function of $\Delta V_{S-T+}$ and the SAW amplitude. The SAW burst duration is 100 ns. The indicated SAW amplitude values correspond to the values at the output of the microwave source. Each data point is the average of 500 single-shot measurements.

In conclusion, we have demonstrated that the existence of a double dot potential close to the transfer position is responsible for spin mixing occurring during the SAW irradiation. In the geometry of the sample, the double dot is aligned with the direction of the SAW propagation and results in a fast change of the detuning during SAW excitation. To reduce the observed spin depolarization during the transfer procedure, an obvious solution is to engineer a double dot potential perpendicular to the SAW propagation axis. In this situation, no detuning change is expected, spin polarization present in the double dot should be preserved even during SAW excitation and fast two-electron spin manipulation would be possible.

# **References**

27. Davies, J. H., Larkin, I. A. & Sukhorukov, E. V. Modeling the patterned two-dimensional electron gas: Electrostatics. *Journal of Applied Physics,* **77**, 4504-4512 (1995).

28. Merkulov, I. A., Efros, Al. L. & Rosen, M. Electron spin relaxation by nuclei in semiconductor quantum dots. *Phys. Rev. B,* **65**, 205309 (2002).

29. Jozsa, R. Fidelity for mixed quantum states. *Journal of modern optics*, *41*(12), 2315-2323. (1994).